\documentclass[11pt]{article}
\usepackage{epstopdf}                                                          
\usepackage[a4paper,hmarginratio=1:1,vmarginratio=2:3,totalwidth=15.2cm,totalheight=22.77cm]{geometry}
\usepackage{bm,epstopdf,epsfig,amsmath,amssymb,
amsfonts,colordvi,wrapfig,comment,cancel,verbatim,slashed}
\usepackage{graphicx,graphics}
\usepackage[font=md,captionskip=8pt]{subfig}
\usepackage[usenames,dvipsnames]{color}
\usepackage[noadjust]{cite}
\usepackage{xcolor} 
\usepackage[utf8]{inputenc}
\usepackage{setspace}

\usepackage[utf8]{inputenc}

\newcommand{\w}{\omega}  
\newcommand{\Om}{\textrm{w}}

\newcommand{\tGamma}{\tilde\Gamma}
\allowdisplaybreaks

\newcommand{\be}{\begin{equation}}
\newcommand{\ee}{\end{equation}}
\newcommand{\bea}{\begin{eqnarray}}
\newcommand{\eea}{\end{eqnarray}}
                
\newcommand{\ra}{\rightarrow}

\newcommand{\baa}{\begin{array}}
\newcommand{\eaa}{\end{array}}

\long\def\symbolfootnote[#1]#2{\begingroup
\def\thefootnote{\fnsymbol{footnote}}\footnote[#1]{#2}\endgroup}

\begin{document} 
\begin{flushright}
\end{flushright}

\thispagestyle{empty}
\vspace{3.cm}
\begin{center}

  {\Large \bf  Weyl quadratic gravity as a gauge theory    \bigskip

                 and non-metricity vs torsion duality}  
  
 \vspace{1.5cm}
 
 {\bf  C. Condeescu},\,\,  {\bf  D. M. Ghilencea}\,\, and  {\bf  A. Micu}\,\,
 \symbolfootnote[1]{E-mail: ccezar@theory.nipne.ro, dumitru.ghilencea@cern.ch, amicu@theory.nipne.ro}
 
\bigskip\bigskip

{\small Department of Theoretical Physics, National Institute of Physics
\smallskip 

 and  Nuclear Engineering (IFIN), Bucharest, 077125 Romania}
\end{center}

\medskip

\begin{abstract}
We review (non-supersymmetric) gauge theories of four-dimensional space-time
symmetries and their  quadratic action. The only  true gauge theory of such a
symmetry (with  a physical gauge boson) that has an exact geometric interpretation,
generates Einstein gravity in its spontaneously broken phase and is
anomaly-free, is that of  Weyl gauge symmetry (of dilatations). Gauging the full
conformal group does not generate a true gauge theory of physical (dynamical)
associated gauge bosons. Regarding the  Weyl gauge symmetry, it is naturally
realised in Weyl conformal geometry, where it admits two different but
equivalent geometric formulations, of same quadratic action:  one  non-metric but
torsion-free, the other  Weyl gauge-covariant and metric (with respect to a
new differential operator). To clarify the origin of this intriguing result, a
third equivalent formulation  of this  gauge symmetry is constructed using the
  standard, modern approach
 on the tangent space (uplifted to  space-time by the vielbein),  which is metric
but has vectorial torsion. This shows  an interesting duality vectorial non-metricity
vs vectorial torsion of the corresponding   formulations, related by a projective transformation.
We comment  on the physical meaning of these results.
\end{abstract}

\newpage


\section{Motivation}

The principle of gauge symmetries has been remarkably successful in high energy physics.
Here we use it in gauge theories of space-time symmetries such as
the Weyl group (Poincar\'e $\times$ dilatations) and
the conformal group, see \cite{Freedman} for a review. In our view  a realistic
gauge theory with such symmetry should: a) recover Einstein gravity in its (spontaneously)
broken phase, b) have a geometric interpretation (as a theory of gravity) and c) be anomaly-free.

Weyl gauge symmetry (of dilatations) is naturally built
in Weyl conformal geometry \cite{Weyl1,Weyl2} (for a review \cite{Scholz})
and thus it does have a  geometric formulation. The Weyl gauge boson of dilatations ($\w_\mu$)
is dynamical, with a field strength $F_{\mu\nu}$ as the length curvature tensor -
a clear geometric origin. This means that  the length is not integrable,
which means that the geometry is non-metric i.e. there is a non-zero
$\tilde\nabla_\mu g_{\alpha\beta}=-2 \w_\mu g_{\alpha\beta}$.
The Weyl gauge symmetry of the associated (Weyl) {\it quadratic} action of this
gauge theory is spontaneously broken \`a la Stueckelberg to Einstein gravity \cite{Ghilen0}, so
$\w_\mu$ becomes massive and decouples, hence non-metricity
effects are strongly suppressed.
Since the Standard Model (SM) with vanishing Higgs mass parameter is scale invariant,
it is naturally  embedded in Weyl geometry with {\it no} additional degrees
of freedom  \cite{SMW}.
This gauge symmetry can be maintained at quantum level which indicates it is
anomaly-free \cite{DG1}, as required for a consistent (quantum) gauge theory.
Successful inflation is possible \cite{WI3,WI1} being just a gauged version of
Starobinsky inflation \cite{Starobinsky}. Good fits for the galaxies
rotation curves are also found \cite{Harko} and associated black hole solutions and
  physics were studied in \cite{Harko2}. All this suggests that Weyl gauge
symmetry with its underlying Weyl conformal geometry are the fundamental
symmetry and geometry beyond the SM and Einstein gravity.

One can also gauge the full conformal group, in which case one obtains
conformal gravity \cite{Kaku} (for a review \cite{Freedman}). However,
in this case the gauge boson of special conformal transformations
$f_{\mu}^a$ is just  an  auxiliary field absent in the final action.
Neither $f_\mu^a$ nor $\w_\mu$ are then dynamical (i.e. physical), hence this  is not a true
gauge theory of the conformal group, in the  high energy theory sense. 
Finally, gauging the Poincar\'e group will generate an action with
an infinite series of higher derivative terms, for which we see  little
motivation.

Returning to Weyl gauge symmetry, it admits \cite{DG1} two equivalent
geometric formulations in Weyl geometry: one is {\it non-metric} but {\it torsion-free}, the other  is
manifestly {\it Weyl gauge covariant and metric} with respect to a new differential operator
$(\hat\nabla)$. This intriguing result requires further investigation and
this is the main motivation of this work.
To this purpose we  construct a gauge theory of dilatations
in a standard tangent space-time approach uplifted to space-time by the
vielbein; this is shown to generate {\it exactly} the Weyl quadratic action associated to Weyl geometry.
This gives  a third equivalent formulation, {\it metric} but {\it with  torsion}, showing a
duality (equivalence) vectorial non-metricity vs vectorial  torsion. All three formulations are
equally good, equivalent descriptions of Weyl quadratic gravity  with this gauge symmetry.
We comment briefly  on some physical aspects of this duality.

\section{Weyl gauge symmetry and geometry: equivalent pictures} 

\subsection{Non-metric formulation}
\label{f1}

Let us discuss Weyl gauge symmetry in its formulation in Weyl geometry\footnote{Our conventions
   \cite{Shapiro}: $g_{\mu\nu}$ with  $(+,-,-,-)$, $g=\vert \det g_{\mu\nu}\vert$.
  To restore the gauge coupling $\alpha$ of dilatations,  rescale $\w_\mu \ra \w_\mu \alpha$.
  For $g_{\mu\nu}$ of  charge $q$  rescale  $\Sigma\!\ra\! \Sigma^{q/2}$.
 We work in  $d\!=\!4\!-\!2\epsilon$, as needed at quantum level.}.
By definition, Weyl geometry is given by equivalence classes 
$(g_{\alpha\beta}, \w_\mu$) of the metric ($g_{\alpha\beta}$)
and the Weyl gauge  field ($\w_\mu$), which in  $d=4-2\epsilon$ dimensions
are related by the transformations below, in the absence
(a) and presence (b) of  scalars ($\phi$) and fermions ($\psi$)

\begin{equation}
  \label{WGS}
  \begin{aligned}
    (a) &\quad  g_{\mu\nu}^\prime=\Sigma^2  \,g_{\mu\nu},\qquad
    \w_\mu'=\w_\mu - \partial_\mu\ln\Sigma,  \qquad
    \sqrt{g'}=\Sigma^{2 d} \sqrt{g}, \\[4pt]
    (b) &\quad \phi' = \Sigma^{q_\phi} \phi, \quad
    \qquad \psi'=\Sigma^{q_\psi}\,\psi,   
  \end{aligned}
\end{equation}
Without loss of generality, for $g_{\mu\nu}$ we set a Weyl charge
$q=2$, then $q_\phi=-(d-2)/2$ and $q_\psi=-(d-1)/2$
as dictated by their canonical kinetic terms.  This defines the
{\it Weyl gauge symmetry} or gauged dilatations symmetry. This should
be distinguished from what is generically called ``Weyl symmetry''
where there is no gauge field.  By definition Weyl geometry is {\it
  non-metric} i.e. $\tilde\nabla_\mu g_{\nu \rho}\not=0$, with:
%
\begin{equation}
  \label{tildenabla}
  (\tilde\nabla_\lambda +2 \, \w_\lambda) g_{\mu\nu}=0,
  \qquad \textrm{where}\qquad
  \tilde\nabla_\lambda g_{\mu\nu}
  =\partial_\lambda g_{\mu\nu}
  -\tilde\Gamma^\rho_{\lambda \mu} g_{\rho\nu}
  -\tilde \Gamma^\rho_{\lambda \nu} g_{\mu \rho}.
\end{equation}

\smallskip\noindent
The Weyl connection $\tilde\Gamma_{\mu\nu}^\rho$ is
found from  (\ref{tildenabla}). In this non-metric formulation
of Weyl geometry one assumes
  a  symmetric connection (i.e. no torsion)
$\tilde \Gamma_{\mu\nu}^\rho\!=\!\tilde\Gamma_{\nu\mu}^\rho$, giving a solution
\begin{equation}
  \label{tildeGamma}
  \tilde \Gamma_{\mu\nu}^\rho
  =\mathring\Gamma_{\mu\nu}^\rho  +
  \big[\delta_\mu^\rho \w_\nu +
  \delta_\nu^\rho \w_\mu-g_{\mu\nu} \w^\rho\big],
\end{equation}
with $\mathring\Gamma_{\mu\nu}^\lambda$ the Levi-Civita (LC) connection.
The Riemann curvature tensor in Weyl geometry  associated to this
connection is defined as in a Riemannian case, but now in terms
of  $(\tilde\Gamma )$:
\smallskip

\begin{equation}
  \label{Rie}
  \tilde R^\rho{}_\sigma{}_{\mu \nu} = \partial_\mu \tilde \Gamma^\rho_{\nu \sigma}
  - \partial_\nu \tilde \Gamma^\rho_{\mu \sigma} + \tilde \Gamma^\rho_{\mu \tau}
  \tilde \Gamma^\tau_{\nu \sigma} - \tilde \Gamma^\rho_{\nu \tau}
  \tilde \Gamma^\tau_{\mu \sigma},
\end{equation}

\medskip\noindent
$ \tilde R^\rho{}_\sigma{}_{\mu \nu} $
can be expressed in terms of $\w_\mu$, for technical details see
Appendix A in \cite{DG1}. From eq.~(\ref{Rie}) one finds 
the expressions of the Ricci tensor  $\tilde R_{\mu \nu}$ and
scalar  $\tilde R$ in Weyl geometry
\begin{align}
  &\tilde R_{\mu\nu}=  \tilde R^\rho{}_{\mu\rho\nu}=    
    \mathring R_{\mu\nu} +
    \frac{d}{2} F_{\mu\nu}-(d-2)\mathring\nabla_{(\mu} \w_{\nu)}
    - g_{\mu\nu}\mathring \nabla_\lambda\w^\lambda +
    (d-2) (\w_\mu\w_\nu -g_{\mu\nu} \w_\lambda\w^\lambda), \label{Ri}
  \\[5pt]
  & \tilde R = g^{\mu\nu}\tilde R_{\mu\nu}=\mathring R-2 (d-1)\,
    \mathring\nabla_\mu \w^\mu  -(d-1) (d-2) \, \w_\mu \w^\mu,\label{Ri2}
\end{align}

\medskip\noindent
 $\mathring R_{\mu\nu}$, $\mathring R$ are
the Ricci tensor and scalar in a Riemannian case, respectively,
$\mathring\nabla$ is  the covariant derivative of Riemannian geometry
(with LC connection); $\mathring\nabla_{(\mu} \w_{\nu)}\equiv (1/2)
(\mathring\nabla_\mu\w_\nu+\mathring\nabla_\nu\w_\mu)$.
While $ \tilde R^\rho{}_\sigma{}_{\mu \nu} $, $\tilde R_{\mu\nu}$ are
invariant since $\tGamma$ is, $\tilde R$ transforms
covariantly under (\ref{WGS}), like $g^{\mu\nu}$.

The Weyl tensor in Weyl geometry ($\tilde C_{\mu\nu\rho\sigma}$) associated
to $\tilde R_{\mu\nu\rho\sigma}$ is related to the  Riemannian
$\mathring C_{\mu\nu\rho\sigma}$ \cite{DG1}
\begin{equation}
  \label{WC}
  \tilde C_{\mu\nu\rho\sigma}^2=\mathring{C}_{\mu\nu\rho\sigma}^2
  + (d^2-2d +4)/(d-2) \,F_{\mu\nu}^2.
\end{equation}

\medskip\noindent
In Weyl geometry there also exists a so-called  length curvature tensor 
$ F_{\mu\nu}=\tilde \nabla_\mu \w_\nu 
-\tilde \nabla_\nu \w_\mu
=\partial_\mu\w_\nu -\partial_\nu \w_\mu,$
which is interpreted as  the field strength of $\w_\mu$, 
where we used that $\tilde\Gamma$ is symmetric and $\tilde \nabla_\mu\w_\nu=
\partial_\mu\w_\nu-\tilde \Gamma_{\mu \nu}^\rho \w_\rho$.
This ends our  geometric  definitions.

With this information, the  {\it most general\,}  Lagrangian of Weyl
quadratic gravity associated to Weyl geometry
in the absence of matter can be written as  \cite{Weyl2}
\bea\label{si}
S=\int d^4x \sqrt{g}    \,\Big\{ a_0\tilde R^2+ b_0 \tilde F_{\mu\nu}^2+
c_0 \tilde C_{\mu\nu\rho\sigma}^2 +d_0 \tilde G\Big\},
\eea
where $a_0$, $b_0$, $c_0$, $d_0$ are constants and $\tilde G$ is  the
Chern-Euler-Gauss-Bonnet term  (hereafter called Euler term) which
is a total derivative (only) for $d=4$; its  expression in $d$ dimensions
is found in \cite{DG1}, eq.(A-14). {\it No other independent terms are allowed in $S$ by
  the symmetry!}

Each term in $S$ is  separately Weyl gauge invariant, as one can easily check.
Since the theory is {\it non-metric}, 
in applications  one is forced to use the (metric) Riemannian formulation obtained from
$S$ by using  relations (\ref{Ri}), (\ref{Ri2}), (\ref{WC}) to curvature tensors
and scalar of  Riemannian geometry. For more technical  details see
Appendix~A in \cite{DG1}.

As discussed extensively  in \cite{Ghilen0}, \cite{SMW},  the gauge theory of
action $S$ has spontaneous breaking \`a la Stueckelberg to Einstein gravity and a small
cosmological constant, after {\it dynamical} $\w_\mu$ becomes massive and decouples
after ``eating'' the dilaton $\ln\phi$; here $\phi$ is the scalar
field propagated by the (geometric)  $\tilde R^2$ term in the action.
Hence, Einstein gravity is just a ``low-energy" effective theory obtained in the
broken phase of action (\ref{si}) and this breaking takes place in the absence of matter.
Mass generation (Planck mass, cosmological constant, $m_\omega$)  has {\it geometric origin},
  being proportional to $\langle\phi\rangle$,  and is also related to a non-vanishing
  (geometric) length-curvature tensor, $F_{\mu\nu}\not=0$ \cite{DG}.
In the presence of the SM,  this mechanism receives corrections from the Higgs
itself, see Section 2.5 in \cite{SMW}, (also \cite{SMW2}) where the phenomenology of
  SM embedded in  Weyl geometry was studied in detail.
  Other phenomenological aspects of action $S$ such as successful inflation
  were discussed in \cite{WI3,WI1} together with interesting implications for dark matter
  \cite{Harko} and black hole physics \cite{Harko2}.

\subsection{Weyl gauge-covariant formulation}\label{f2}

For a gauge theory one would actually like to have manifest Weyl
gauge-covariance.  The gauge theory formulation in Section~\ref{f1} is
not entirely satisfactory because it is not manifestly covariant, as
one can easily see: the partial derivative $\partial_\mu$ in
$\tilde \nabla_\mu$ when acting on the (geometric) tensors like
$\tilde R_{\mu\nu}$, etc, or on scalar $\tilde R$, is not Weyl
gauge-covariant.  The explanation is that one should account for the
effect of their Weyl charges in the derivative
acting on them, etc.  A related issue is that the geometry is not metric
($\tilde\nabla_\mu g_{\nu \rho}\!\not=\!0$) making calculations
difficult and forcing one to go to a Riemannian picture.

The non-metricity and the absence of manifest Weyl gauge covariance in
the previous geometric formulation can be addressed and solved
simultaneously.  Since
$(\tilde\nabla_\lambda+q \w_\lambda) g_{\mu\nu}=0$, where $q=2$ is the
charge of $g_{\alpha\beta}$, this suggests that for any  tensor
$T$, including $g_{\mu\nu}$, of Weyl charge\footnote{The charge $q_T$ is in principle arbitrary. For the
  objects used in this paper they are given on page
  \pageref{charges}.}
$q_T$
($T'=\Sigma^{q_T} T$) one should introduce a new differential operator
$\hat\nabla$ (replacing $\tilde\nabla$)
\medskip
\begin{equation}
  \hat\nabla_\lambda T \equiv (\tilde \nabla_\lambda+q_T  \, \w_\lambda) \, T
\end{equation}

\medskip\noindent
This new operator transforms covariantly under (\ref{WGS}), as seen by using that
$\tilde \Gamma$ is invariant: $\hat\nabla_\mu^\prime T^\prime=\Sigma^{q_T} \hat\nabla_\mu
T$. The theory is then metric with respect to the  new operator: $\hat\nabla_\mu g_{\alpha\beta}=0$.

For reasons that become clear shortly, we also define a new Riemannian and Ricci
tensors and Ricci scalar of Weyl geometry  \cite{DG1,Tann,Jia}
\medskip
\be\label{nat}
\hat R_{\mu\nu\rho\sigma}=\tilde R_{\mu\nu\rho\sigma}-g_{\mu\nu} \hat F_{\rho\sigma},
\quad
\hat R_{\nu\sigma}=\tilde R_{\nu\sigma}-  \hat F_{\nu\sigma},
\quad
\hat R=\tilde R.
\ee

\medskip\noindent
with  $\hat F_{\mu\nu}= F_{\mu\nu}=\partial_\mu \w_\nu-\partial_\nu \w_\mu$. Note also
that $\hat R_{\mu\nu}-\hat R_{\nu\mu}=(d-2) F_{\mu\nu}$, relevant later.
With (\ref{Rie}), (\ref{Ri})
 one easily writes these curvatures in terms of their Riemannian
 counterparts.

 One benefit of the new ``hat'' basis is that the new Weyl tensor  $\hat C_{\mu\nu\rho\sigma}$
 associated  to $\hat R_{\mu\nu\rho\sigma}$   and Euler terms become \cite{DG1} (Section 3.1)
\bea
\hat C_{\mu\nu\rho\sigma}=\mathring C_{\mu\nu\rho\sigma},
\qquad
\hat G= \hat R_{\mu\nu\rho\sigma} \hat R^{\rho\sigma\mu\nu}
- 4 \hat R_{\mu\nu} \hat R^{\nu\mu} +\hat R^2.
\eea

\medskip\noindent
The new Weyl tensor  is identical to that in Riemannian geometry, while
$\hat G$ is $\tilde G$ of previous section but in the ``hat basis'' and
is a  generalisation to Weyl geometry of the  Euler term.

A second important benefit is the  Weyl gauge covariance under transformation (\ref{WGS})
\medskip
\bea\label{WGS3}
 X^\prime=\Sigma^{-4} X, \qquad
 X
 =\hat R_{\mu\nu\rho\sigma}^2, \, \,\hat R_{\mu\nu}^2,\,\,\hat R^2,
 \,\, \hat C_{\mu\nu\rho\sigma}^2,\,\, \hat G,
\,\, \hat F_{\mu\nu}^2.
\eea
 \bea\label{wq}
\hat\nabla_\mu' \hat R^\prime=\Sigma^{-2}\hat \nabla_\mu \hat R,\qquad
\hat\nabla^\prime_\mu\hat\nabla^{\prime\,\mu} \hat R'
=\Sigma^{-4} \hat \nabla_\mu\hat\nabla^\mu \hat R, \quad 
\hat\nabla_\rho' \hat R_{\mu\nu}'=\hat\nabla_\rho \hat R_{\mu\nu},\,\,\,
\text{etc.}
\eea
%
Unlike its Riemannian version, the Euler term $\hat G$  is now
Weyl covariant in arbitrary $d$ dimensions (just like  $\hat C_{\mu\nu\rho\sigma}^2$) which
is very important for maintaining  this symmetry at quantum level and avoiding
the Weyl anomaly \cite{DG1}.
With this information,  action (\ref{si}) becomes 
\bea\label{SS}
S=\int d^4x  \sqrt{g}\,\Big\{ a_0\hat R^2+ b_0 \hat F_{\mu\nu}^2+
c_0 \hat C_{\mu\nu\rho\sigma}^2 +d_0 \hat G\Big\}.
\eea
%
up to a redefinition of $b_0$. 
Each term in $S$  is again separately invariant under (\ref{WGS}) for $d=4$.

The  Weyl covariance of $\hat R$  enables us to maintain  Weyl gauge symmetry
also in $d=4-2\epsilon$ dimensions by a natural ``geometric'' analytical continuation
\bea\label{Sf}
S=\int d^d x  \sqrt{g}\,\Big\{ a_0\hat R^2+ b_0 \hat F_{\mu\nu}^2+
c_0 \hat C_{\mu\nu\rho\sigma}^2 +d_0 \hat G\Big\} \, \hat R^{2 (d-4)/4}.
\eea
Quantum calculations can now be done \cite{DG1} in this metric-like,  Weyl gauge covariant
picture\footnote{This Weyl invariant regularisation
  implicitly assumes $\tilde R\not=0$, which is verified a-posteriori \cite{DG1}.}.

To conclude, with respect to the new $\hat\nabla$ operator
we simultaneously have a  {\it metric-like} formulation and
a  Weyl gauge-covariant  description of {\it geometric} operators
(curvature tensors/scalar)  {\it and}  of  their derivatives, as in any gauge theory.
Action (\ref{SS}) is equivalent  to (\ref{si}) up to a re-definition of $b_0$,
so it gives the same physics.
We thus presented a {\it manifestly  covariant}, metric formulation of Weyl geometry 
as a gauge theory of space-time dilatations. Quantum calculations can now be done
directly in this (metric) formulation of Weyl geometry \cite{DG1} using (\ref{Sf})
while keeping a  manifest Weyl gauge symmetry in $d$ dimensions for each term in the action;
in this way one shows that $S$ of (\ref{Sf}) is  anomaly-free \cite{DG1}.

\subsection{Tangent space formulation has torsion}\label{f3}

In the previous sections we presented the  Weyl gauge symmetry from its 
realisation in Weyl geometry, using  a geometric approach that lead to
two equivalent formulations.
This equivalence  demands  some clarification in  the  modern gauge theory approach.
We do this by constructing the gauge theory of the Weyl group on the tangent
space and uplifting it to space-time by the vielbein, see
\cite{Freedman,Charap}.

The Weyl group is a subgroup of the conformal group which consists of
the Poincar\'e group $\times$ dilatations (it does not include special
conformal transformations). The gauge algebra is
\begin{equation}
\begin{split}
  \label{gauge-algebra}
  &  [P_{a}, M_{bc}] =  \eta_{a b} P_{c} - \eta_{ac}P_b, \qquad   [D,P_a] = P_a,
  \qquad [P_a, P_b] = 0, \qquad    [D, M_{ab}] = 0, \\[5pt]
& [M_{ab}, M_{cd}] = \eta_{a c}M_{d b} - \eta_{bc}M_{da} - \eta_{ad} M_{cb} + \eta_{bd} M_{ca},
\end{split}
\end{equation}

\medskip\noindent
where $\eta_{ab}$ is the Minkowski metric and $a, b, \ldots $ denote tangent
space indices.  $P_a$,  $M_{ab}$ and $D$ are the generators of translations,
Lorentz transformations (rotations) and  dilatations, respectively.  
Their associated gauge fields are the vielbein $e_\mu^a$, spin connection $\Om_\mu{}^{ab}$
and Weyl boson $\w_\mu$, respectively. The  corresponding structure constants can
be read from  the  Lie algebra
$[T_A,T_B]= f_{AB}{}^C T_C$, where $T_A$ stands for $P_a$, $M_{ab}$, $D$.
The  gauge curvature  $R_{\mu \nu}^A$ of the  gauge field $B_\mu^A$ is
 $ R_{\mu \nu}^A = 2 \partial_{[\mu} B_{\nu]}^A + B_\mu^B B_\nu^C f_{BC}{}^A$, 
(here $x_{[\mu} y_{\nu]}\equiv (1/2)(x_\mu y_\nu - x_\nu y_\mu)$).

With the structure constants from (\ref{gauge-algebra}) we find
the field strength of local translations, rotations and dilatations
\begin{align}
  R_{\mu \nu} (P^a) &= 2 D_{[\mu} e_{\nu]}^a +2 \w_{[\mu} e_{\nu]}^a,
                      \label{translations}\\
R_{\mu\nu}(M^{ab}) & =  \partial_\mu \Om_\nu{}^{ab} -
  \partial_\nu \Om_\mu{}^{ab} + \Om_\mu{}^a{}_c \Om_\nu{}^{cb}
  - \Om_\nu{}^a{}_c \Om_\mu{}^{cb} \equiv R^{ab}{}_{\mu \nu} \, , \\
R_{\mu \nu}(D) & =\partial_\mu \w_\nu - \partial_\nu \w_\mu \equiv  F_{\mu \nu} \, ,
 \end{align}
 where
 \bea
 D_\mu e_\nu^a = \partial_\mu e_\nu^a + \Om_\mu{}^a{}_b e_\nu^b,
 \eea
 is the Lorentz covariant derivative.
$F_{\mu \nu}$ denotes the field strength of the Weyl gauge field of  dilatation
$\w_\mu$,  and $R^{a}{}_{b\mu \nu}$ is the usual two-form
curvature tensor defined from the commutator of the tangent space
(Lorentz) covariant derivatives
\begin{equation}
R^a{}_{b\mu \nu} : = e_b^\sigma [D_\mu, D_\nu] e_\sigma^a.
\label{bla-1}
\end{equation}

\medskip\noindent
Under a general (infinitesimal) gauge transformation
$\delta_\epsilon \equiv \epsilon^A T_A = \xi^a P_a + (1/2)
\lambda^{ab} M_{ab} + \lambda_D D$ the gauge field change as
$\delta_\epsilon B_\mu^A = - \partial_\mu \epsilon^A + \epsilon^B
B_\mu^C f_{BC}{}^A$, while the curvatures transform covariantly
$\delta_\epsilon R_{\mu \nu}^A =\epsilon^B R_{\mu \nu}^C f_{BC}{}^A$.
For the case at hand, considering only dilatations (i.e. setting to zero all
 gauge parameters except $\lambda_D$) we find
\medskip
\begin{equation}
  \delta_\epsilon e_\mu^a  = \lambda_D e_\mu^a,\qquad
  \delta_\epsilon \Om_{\mu}{}^{ab}  =0, \qquad
\delta_\epsilon \w_\mu  = -\partial_\mu \lambda_D, \label{var-3}
\end{equation}
and
\begin{equation}
  \delta_\epsilon R_{\mu \nu}(P^a) =\lambda_D R_{\mu \nu}(P^a), \quad
\delta_\epsilon R_{\mu \nu}(M^{ab}) =0, \quad
\delta_\epsilon R_{\mu \nu}(D)  = 0.
\label{t2}\end{equation}

\medskip\noindent
Notice that eq.(\ref{var-3}) is an infinitesimal version
of (\ref{WGS}) of Weyl geometry, with $\Sigma=\exp(\lambda_D)$.

Let us mention the particular case of gauging the Poincar\'e symmetry
recovered from the above formulae by setting $\w_\mu=0$. The
diffeomorphism invariance of the theory is then implemented by the
constraint $R_{\mu\nu}(P^a)=0$ which in the Poincar\'e case gives
$D_{[\mu} e_{\nu]}^a=0$. This is just the first Cartan
  structure equation without torsion which gives the well-known result
  for the spin-connection
$\mathring{\Om}_\mu{}^{ab} = 2 e^{\nu [a} \partial_{[\mu}
e_{\nu]}{}^{b]} - e^{\nu [a} e^{b] \sigma} e_{\mu c} \partial_\nu
e_\sigma{}^c$.

Compared to the Poincar\'e  case,  $R_{\mu\nu}(P^a)$ 
in eq.\eqref{translations}  contains now
an extra term due to $\w_\mu$. This term can be  interpreted as torsion
in the first Cartan structure equation
\begin{equation}
D_{[\mu} e_{\nu]}^a = - 2 \w_{[\mu} e_{\nu]}^a \equiv T_{\mu\nu}{}^a
\label{cartan-torsion}
\end{equation} 

\medskip\noindent
As a result,  the curvature constraint  $R_{\mu\nu}(P^a)=0$ gives
a Weyl spin connection $\Om_\mu{}^{ab}$
\medskip
\begin{equation}
\Om_\mu{}^{ab} = \mathring{\Om}_\mu{}^{ab} +2 e_\mu^{[a} e^{b]\nu} \w_\nu.
\label{spin-invariant}
\end{equation}

\medskip\noindent
It is important to note that the constraint  $R_{\mu \nu}(P^a)=0 $
is invariant under dilatations, see (\ref{t2}). Since the
original spin-connection is also invariant, see (\ref{var-3}), this
guarantees that  solution \eqref{spin-invariant} does not
transform under dilatations.
Furthermore, the curvature two-form $R^{a}{}_{b \mu \nu}$ is
also invariant and hence is the correct geometrical object
(together with $F_{\mu \nu}$) for building an invariant action.

The above tangent space formulas can  now be ``uplifted'' to  space-time
with the vielbein. The affine connection
$\Gamma_{\mu \nu}^\rho \equiv e_a^\rho D_\mu e_\nu^a$ corresponding to
$\Om_\mu{}^{ab}$ becomes
\medskip\begin{equation}
  \label{affine-1}
  \Gamma_{\mu \nu}^\rho = \mathring \Gamma_{\mu \nu}^\rho
  +  \delta_\mu^\rho \w_\nu  - g_{\mu\nu} \w^\rho \, ,
\end{equation}  

\medskip\noindent
and is metric compatible $\nabla_\mu g_{\nu \rho} =0$ but now we have torsion
$T_{\mu\nu}{}^\rho \equiv\Gamma^\rho_{\mu \nu}-\Gamma_{\nu\mu}^\rho =2
\delta_{[\mu}^\rho \w_{\nu]}$. 
For a later discussion, notice  that  $\Gamma$ is related  to symmetric $\tilde\Gamma$
of (\ref{tildeGamma}) of the non-metric formulation, by a projective transformation\footnote{See \cite{Sauro:2022hoh} for more on projective transformations in the context of Weyl geometry.} 
\bea
\tilde\Gamma_{\mu\nu}^\rho=\Gamma_{\mu\nu}^\rho +\delta_\nu^\rho \w_\mu.
\eea

\medskip\noindent
Further,  the  Riemann tensor associated to $\Gamma$ is the uplifted version of eq.\eqref{bla-1}
\medskip
\begin{equation}
\label{riemann-2}
  R^\rho{}_\sigma{}_{\mu \nu} = \partial_\mu \Gamma^\rho_{\nu \sigma}
  - \partial_\nu \Gamma^\rho_{\mu \sigma} + \Gamma^\rho_{\mu \tau}
  \Gamma^\tau_{\nu \sigma} - \Gamma^\rho_{\nu \tau} \Gamma^\tau_{\mu \sigma},
\end{equation}

\medskip\noindent
and it is antisymmetric in both the first and last pair of indices; however, it is not symmetric
in the exchange of the first pair with the last pair. One finds
\medskip
\begin{align}
  R_{ \rho \sigma \mu \nu } & = \mathring{R}_{ \rho \sigma \mu \nu }
                              + \left[g_{\mu \sigma} \mathring{\nabla}_\nu \w_\rho
                              - g_{\mu \rho} \mathring{\nabla}_\nu  \w_\sigma
                              + g_{\nu \rho}  \mathring{\nabla}_\mu
\w_\sigma - g_{\nu \sigma}\mathring{\nabla}_\mu \w_\rho  \right] \nonumber\\
                            &+ \w^2(g_{\mu \sigma} g_{\nu \rho} - g_{\mu \rho} g_{\nu \sigma})
                              + \w_\mu(\w_\rho g_{\nu \sigma} - \w_\sigma g_{\nu \rho})
                              + \w_\nu(\w_\sigma g_{\mu \rho} - \w_\rho g_{\mu \sigma})\, ,
                              \label{riem-1}\\[5pt]
  R_{\mu \nu} & = \mathring{R}_{\mu \nu} - (d-2) \mathring{\nabla}_\nu \w_\mu- g_{\mu \nu}
                \mathring{\nabla}_\alpha \w^\alpha+ (d-2)\w_\mu \w_\nu - (d-2)g_{\mu \nu}\w^\alpha \w_\alpha
       \label{riem-2} \\[5pt]
R & = \mathring{R} - 2(d-1)\, \mathring{\nabla}_\mu \w^\mu - (d-1)(d-2)\, \w_\mu \w^\mu \, .\label{riem-3}
\end{align}

\medskip
Remarkably,  the expressions for $R_{\rho\sigma\mu\nu}$, $R_{\mu\nu}$ and $R$
are  identical to those in the Weyl covariant formulation
of eq.(\ref{nat}) with replacements (\ref{Rie}), (\ref{Ri}),
and obtained in the ``hat'' basis which is  metric  with respect  to $\hat\nabla_\mu$.
Below we  clarify the origin of this equivalence.

In a true gauge theory we need fully covariant derivative
operators. Therefore we introduce the derivative $\hat D_\mu$ by its
action on a tangent space vector $V^a$ of (arbitrary) Weyl weight
$q_V$
\begin{equation}
\hat D_\mu V^a = \partial_\mu V^a + q_{V}\, \w_\mu\, V^a + \Om_\mu{}^a{}_b\, V^b \, .
\end{equation}  

\medskip\noindent
Since $\hat D_\mu$ coincides with the standard tangent space derivative $D_\mu$
(defined by a spin connection $\Om_\mu{}^{ab}$) when acting on tensors with
zero Weyl weight it is straightforward to see that
$\hat D_\mu$ is  compatible with the metric $\eta_{ab}$ as
$\hat D_\mu \eta_{ab} = D_\mu \eta_{ab} = 0$.

Translating this derivative $\hat D_\mu$ to space time by
\medskip
\begin{equation}
\hat \nabla_\mu V^\nu = e_a^\nu \hat D_\mu V^a \, ,
\end{equation}     

\medskip\noindent
we find precisely $\hat \nabla_\mu$ defined in the previous section.

Consider now a Weyl invariant vector on the tangent space $V^a$. We can write
\begin{equation}
  \hat \nabla_\mu V^\nu = e_a^\nu \hat D_\mu V^a = e_a^\nu D_\mu V^a
  = \nabla_\mu V^\nu \, .
\end{equation}
\medskip
This implies \cite{Cezar}
\medskip
\begin{equation}
  [\hat \nabla_\mu, \hat \nabla_\nu] V^\rho = R^\rho{}_{\sigma \mu \nu} V^\sigma  \, ,
\end{equation}

\medskip\noindent
which shows that the Riemann tensor $\hat R^\rho{}_{\sigma \mu \nu}$
associated to the metric gauge covariant derivative $\hat \nabla_\mu$
is geometrically expressed in terms of a connection with torsion
(see eqs.\eqref{affine-1} and \eqref{riemann-2}). In conclusion
we have shown that we have the identity
\begin{equation}\label{relation}
\hat R^\rho{}_{\sigma \mu \nu} = R^\rho{}_{\sigma \mu \nu} \, ,
\end{equation}

\medskip\noindent
and similar relations for its contractions, as already checked, see text after eq.(\ref{riem-3}).
This also confirms that  the tensors $R^\rho{}_{\sigma \mu \nu}$ and $R_{\mu\nu}$ are
invariant while $R$  transforms covariantly under the  gauged dilatation transformation,
as already seen in Section~\ref{f2}.

In conclusion,  the Weyl gauge-covariant picture of Section~\ref{f2} 
gives  rise to the same curvature tensors/scalar as in the formulation of this section
that is metric, with torsion.

We can  now write the action for the gauge theory of the Weyl group.
It is natural to consider the most general invariant action quadratic
in the curvatures, as in any gauge theory,  with indices contracted
with the metric $g_{\mu \nu}$ or the completely antisymmetric $\epsilon$-density
$\epsilon_{\mu_1 \ldots \mu_d}$ (or their tangent space counterparts).
To derive the general action, one uses  the Weyl charges of various fields
under gauged dilatations, which are:
\medskip
\begin{table}[h]
  \label{charges}
  \centering
  \begin{tabular}{|c|c|c|c|c|c|c|c|c|c|c|c|c|} 
 field & $e_\mu^a$ & $e_a^\mu$ & $g_{\mu\nu}$ & $g^{\mu\nu}$ & $\Om_\mu^{\,\,\,ab}$ & $\sqrt{g}$
    & $R^{a}{}_{b\mu\nu}$ & $R_{\mu\nu}$ & $R$ & $F_{\mu\nu}$
    & $\phi$ & $\psi$
    \\[2mm]
    \hline
    & & & & & & & & & & & &    \\[-3mm]
$q$ &   1 & -1 & 2 & -2 & 0 & $d$  &0 & 0 & -2 & 0
 & $-\frac{d-2}{2}$ & $-\frac{d-1}{2}$ 
  \end{tabular}
\end{table}

\medskip
By analysing the symmetries of the possible terms, one shows that there are
four independent terms in the action, 
$R^2, R_{(\mu \nu)} R^{(\mu \nu)}, R_{\mu \nu \rho \sigma} R^{\mu \nu \rho \sigma}$
and $F_{\mu \nu} F^{\mu \nu}$ or their combinations. In $d=4$ one
can also build topological invariants by using the $\epsilon$-density. We consider the
Euler term\footnote{For a four-dimensional manifold $M$ (compact, orientable, without border) the Euler
  characteristic can be computed from a general {\it metric} connection with the formula $
  \chi (M) = \int_M e(R) = 1/(2\pi)^2 \int_M \text{Pf} (R) = 1/(2 \pi)^2
  \int_M 1/(2!\, 2^2) \epsilon_{abcd} R^{ab} \wedge R^{cd} = 1/(32\pi^2)
  \int d^4x \, \sqrt{g} \left(R^2 - 4R_{\mu \nu} R^{\nu \mu}
    + R_{\mu \nu \rho \sigma} R^{\rho \sigma \mu \nu} \right)
  $.} term $G$, which for a connection with torsion is given by
\begin{equation}
G = R^2 - 4R_{\mu \nu} R^{\nu \mu} + R_{\mu \nu \rho \sigma} R^{\rho \sigma \mu \nu}.
\end{equation} 

\medskip\noindent
Notice the position of the contracted indices which is essential in making $G$ a topological
invariant for a connection with torsion in four dimensions.
In $d$ dimensions, $G$ is no longer a topological invariant but  it is Weyl gauge-covariant
like its counterpart in Section~\ref{f2} to which is actually identical.
A convenient choice of independent quadratic terms each invariant under gauged dilatations
gives the following action (with  constants $a_0,..., d_0$):
\medskip
\begin{equation}
  S = \int d^4x \sqrt{g}\,  \left[a_0\, R^2 + b_0\, F_{\mu \nu}^2
 + c_0   C_{ \rho \sigma \mu \nu}^2  + d_0\, G \right] \, .
\label{action4}
\end{equation}

\medskip
This action is identical (up to a redefinition of couplings $a_0$,..., $d_0$)
to that discussed in the {two}  ``geometric'' formulations of the previous sections.
In $d$ dimensions this action can be continued analytically as in eq.(\ref{Sf}).
In conclusion, gauging the Poincar\'e $\times$ dilatations symmetry
gives rise to the same theory  as in the non-metric or in the Weyl gauge-covariant
formulations, so  we have (in $d$ dimensions) three equivalent formulations
of this symmetry\footnote{There is a special limit of action (\ref{action4}) when
  $\w_\mu$  is ``pure gauge'', so  $F_{\mu\nu}=0$; $\w_\mu$ 
can then  be integrated out, to leave  an action with Weyl symmetry only (no $\w_\mu$ field), see
\cite{DG,SMW2} for an extensive  discussion.}.

\section{Torsion vs non-metricity duality}

So far we found three different formulations of theories with Weyl
gauge symmetry: one in terms of a non-metric connection
$\tilde \Gamma$, one in terms of a (metric) connection with torsion
$\Gamma$ and one fully covariant formulation in terms of the operators
$\hat \nabla_\mu$ and $\hat D_\mu$. In this section we want to analyse
more closely the relation between these formulations.

The non-metric connection ($\tilde\Gamma$) of \eqref{tildeGamma} 
is invariant under  transformations eq.\eqref{WGS},
symmetric in $(\mu, \nu)$ and  thus  torsionless,
but it does not preserve the metric:
$\tilde \nabla_\mu g_{\nu \rho} = - 2 \w_\mu g_{\nu \rho}$.  The spin
connection associated to it can be computed from the
usual formula
\medskip
\begin{equation}
  \label{tildeomega}
  \tilde \Om_\mu{}^a{}_b = - e^\nu_b \tilde \nabla_\mu e_\nu^a
  = \mathring \Om_\mu{}^a{}_b + e_\mu^a e_b^{ \nu} \w_\nu - e_{b \mu} e^{a \nu} \w_\nu+ \delta_b^a \w_\mu \ . 
\end{equation}

\medskip\noindent
The last term in the rhs, symmetric in $(a,b)$,  spoils the
tangent space metricity because
\medskip
\bea
\tilde D_\mu \eta_{ab} = - 2\w_\mu  \eta_{ab}.
\eea

\medskip\noindent
Unlike $\Om_\mu{}^{ab}$ which was Weyl gauge invariant,
$\tilde\Om_\mu{}^{ab}$ transforms like a gauge field (its trace is proportional to
$\w_\mu$). This gives the non-metricity one-form with components
$Q_{\mu ab}= -2 \w_\mu \eta_{ab}$. Similarly, in the tangent space
formulation we had the torsion two-form given by
\eqref{cartan-torsion}. Both formulations comprise additional degrees
of freedom compared to Riemannian geometry. In both cases the extra
degrees of freedom are vectors (in d dimensions) which are identified
with the Weyl gauge boson $\w_\mu$. Therefore we have a special
relation between vectorial non-metricity and vectorial torsion on
which we shall comment later. One can associate a curvature tensor
to $\tilde\Om$  via the commutator
\medskip
\begin{equation}
  \tilde R^{a}{}_{b \mu \nu} :=  e_b^\sigma [\tilde D_\mu, \tilde D_\nu] e_\sigma^a
  =\partial_\mu \tilde \Om_\nu{}^{a}{}_b -
  \partial_\nu \tilde \Om_\mu{}^{a}{}_b + \tilde \Om_\mu{}^a{}_c \tilde \Om_\nu{}^{c}{}_b
  - \tilde \Om_\nu{}^a{}_c \tilde \Om_\mu{}^{c}{}_b \, ,
\end{equation}

\medskip
This corresponds to the usual curvature tensor in the non-metric formulation of Weyl
gravity which gives eq.(\ref{Rie}).
Note that the only symmetry of this tensor is the antisymmetry in the
last two indices.
With this, the curvature tensor, Ricci tensor and scalar in the tangent space and
non-metric formulations are then related
\begin{equation}
  \begin{aligned}
    \label{correspondence}
    R_{\rho \sigma \mu\nu}  &= \tilde R_{\rho \sigma \mu \nu} - g_{\rho \sigma} F_{\mu \nu},
    \qquad
    & R_{\mu \nu} & = \tilde R_{\mu \nu} - F_{\mu \nu},\qquad  &
    R & = \tilde R \, ,
\end{aligned}
\end{equation}

\medskip\noindent
as already noticed in eq.\eqref{nat}, (\ref{relation})  in the Weyl
covariant picture.

We now have a clear description of the transition between the tangent
space formulation with torsion and the (torsion-free) non-metric formulation: the
affine and spin connections of these formulations are related by a projective transformation
\medskip
\begin{equation}
  \begin{aligned}
    \tilde \Om_\mu{}^{ab} &  = \Om_\mu{}^{ab} + \eta^{ab} \w_\mu \, , \\
    \tilde \Gamma_{\mu \nu}^\rho & = \Gamma_{\mu \nu}^\rho + \delta_\nu^\rho \w_\mu
    \, ,
  \end{aligned}
\end{equation}

\medskip\noindent
where the last terms in the rhs of these equations account for the
non-metricity of the lhs connections. With the new (spin) connection $\tilde \Om$,
eq.(\ref{cartan-torsion}) becomes $ \tilde D_{[\mu}   e_{\nu]}^a = 0$,
and thus has zero torsion. Hence the same equation admits two
interpretations, one in terms of torsion and the other in terms of
non-metricity. We thus have a ``dual picture'' and interpretation of vectorial
torsion vs vectorial non-metricity (see \cite{Klemm,Iosifidis} for a related study).

The vielbein postulate can also be written in  different ways,
depending on which affine and spin connections one is using.
Indeed, we have the following equivalent equations
\medskip
\begin{align}
  \nabla_\mu e_\nu^a + \Om_\mu{}^a{}_b e_\nu^b &= 0 \, ,\label{a1}
  \\
  \tilde D_\mu e_\nu^a - e_\rho^a \tilde \Gamma_{\mu \nu}^\rho & = 0 \, ,
              \label{a2}\\
  (\tilde \nabla_\mu + \w_\mu) \, e_\nu^a + \Om_\mu{}^a{}_b e_\nu^b & = 0 \, .
      \label{a3}
\end{align} 

\medskip
Eq.(\ref{a1}) reflects the choice of working with the metric affine
connection $\Gamma_{\mu\nu}^\rho$ of eq.\eqref{affine-1} with torsion,
and the invariant (and metric) spin connection\footnote{The spin
  connection is invariant because in this case
  $\nabla_\mu e_\nu^a = \hat \nabla_\mu e_\nu^a $ and hence the first
  term is covariant.}  of eq.\eqref{spin-invariant}, as in
Section~\ref{f3}.

Eq.(\ref{a2}) implies that one is choosing an invariant non-metric
affine connection \eqref{tildeGamma} paired with a non-invariant (and
non-metric) spin connection \eqref{tildeomega} which, however,
covariantises the corresponding tangent space derivative
$\tilde D_\mu$ when acting on the vielbein (since
$\tilde D_\mu e_\nu^a = \hat D_\mu e_\nu^a$).  Therefore,
eqs.(\ref{a1}) and (\ref{a2}) pair (non-)metricity in the space-time
with (non-) metricity on the tangent space, respectively.
This was used in Section~\ref{f1}.

A mixed choice is also possible. Indeed, in eq.(\ref{a3}), because
$\tilde \nabla_\mu e_\nu^a$ is not covariant with respect to
dilatations, one adds a further covariantisation
$(\tilde \nabla_\mu + \w_\mu) e_\nu^a= \hat \nabla_\mu e_\nu^a$.  This
is the choice that corresponds to the Weyl covariant picture in
Section~\ref{f2}, with both the affine and spin connections invariant
and seems suitable for physical applications. This case pairs a
non-metric connection in space-time with a metric spin connection on
the tangent space.

There is an additional interesting aspect of the duality we found
(covariant) non-metric versus torsion formulations. It is well-known
that connections with torsion preserve the norm of vectors under
parallel transport. In agreement with our equivalence of formulations,
and contrary to a long-held (wrong) view, the (torsion-free)
non-metric formulation of Weyl geometry also preserves the norm of the
vectors under their parallel transport along a curve. This result
applies provided that 1) vectors are Weyl invariant in the tangent
space (i.e. vanishing charge in tangent space $q_v=0$), and 2) their
parallel transport preserves the Weyl gauge covariance, as demanded in
a gauge theory, something missed by the long-held view.  This result
is shown in eqs.(B-8) to (B-13) in \cite{DG}.  This is consistent with
the above equivalence of the formulations of Weyl geometry as a gauge
theory of gravity.

More generally, for vectors of arbitrary
tangent-space charge ($q_v\not=0$), parallel transport is again
physically meaningful only if Weyl gauge covariance is maintained, so
the gauge covariant derivative (i.e. $\hat \nabla$) is used; since
this operator is metric compatible, the norm changes only by the
charge of the tangent space vector.
To detail, consider the infinitesimal covariant parallel transport
of a vector $v$ i.e.  $ dx^\mu \hat\nabla_\mu v^\nu = 0$, then by the metricity of
the gauge covariant  derivative ($\hat\nabla_\mu g_{\alpha\beta}=0$), one has that the norm
is covariantly constant $ dx^\mu \hat \nabla_\mu |v|^2 = 0$.  This  implies
the following variation  $d\, |v|^2 = -2 q_v \w_\mu |v|^2dx^\mu $ with $q_v$ the tangent
space charge; if $q_v=0$ we re-obtain the  norm is invariant.
This result is identical in the metric formulation with torsion (using $\hat\nabla$)
or non-metric covariant formulation \cite{DG} (eq.(B-12)).

In conclusion, for a description of Weyl gauge symmetry all three
formulations are equally good. Weyl gauge symmetry does not 
prefer one connection or the other, although, from a high energy theory viewpoint,
the Weyl-covariant  formulation may be preferable. The above equivalence of the
three formulations of the quadratic  gravity, as a gauge theory associated to Weyl geometry,
is specific  for the vectorial non-metricity of Weyl geometry (and vectorial torsion),
but the situation changes in more  general cases \cite{Cezar}.
This is easily understood, because torsion and
non-metricity have in general a different physical meaning.
This distinction is more intuitive in solid state physics,
see section 4.4 in \cite{DG}.  Consider a 3D crystalline
structure: defects of dimension $d=0$ known as point defects (missing
atoms, extra atoms, etc) that destroy the local notion of length are
naturally associated with non-metricity. Torsion is associated with
defects of dimension $d=1$ known as dislocations of the
lattice. Hence, there is a clear difference between torsion and
non-metricity.  Then why is there no such difference apparent in our
study above?

To understand this, note that we only considered vectorial
non-metricity and vectorial torsion, that lead to the dual, equivalent
interpretations.  This is because both torsion and non-metricity have
a vector component under $so(4)$ algebra decomposition, which is
``tested'' here.  But torsion and non-metricity tensors have
additional degrees of freedom beyond this vector component that do
distinguish between these two tensors both mathematically and
physically. In other words, the equivalent dual interpretation
discussed here will fail beyond the vectorial non-metricity/torsion and
then the physical aspects of non-metricity and torsion are indeed
different in a general case \cite{Cezar}.

So far we discussed only gauged dilatations. The general result by
Coleman-Mandula \cite{Coleman} allows us to have the conformal group
as the maximal space-time symmetry. In addition to the Weyl group, the
conformal group includes special conformal transformations. Using
these transformations we can always set to zero the gauge field
$\w_\mu$ of dilatations\footnote{It is for  this reason that one can construct
  Poincar\'e gravity/supergravity   as gauged fixed theories with
  conformal/superconformal symmetry.} \cite{Freedman}. Moreover, at quadratic order
in curvatures, no kinetic term for the gauge field of special
conformal transformations can be written, so the
corresponding gauge field is not dynamical (physical), either \cite{Kaku}.
Thus, in this case we cannot talk about a true gauge
theory (in the same way the  electroweak theory without  kinetic terms for
the gauge bosons $W^\pm$, $Z$ cannot be regarded as a gauge theory of weak interactions).
Therefore, only gauged dilatations  give a true (and anomaly-free) gauge theory
of a four-dimensional space-time symmetry of the action.  In this case, Weyl geometry
seems the natural  underlying geometry that realises this symmetry, even in the
absence of matter. It may  actually be the unique geometry to do so in a realistic way,
given the equivalent dual formulation we found, as discussed in  \cite{Cezar}.

So far our analysis  did not discuss the effect of adding matter fields.
It is easy to see that our results remain valid when the SM is embedded
in Weyl geometry. First, the SM gauge sector is invariant under (\ref{WGS}) while
the fermions Dirac action  is identical to that in Riemannian
geometry and is invariant under (\ref{WGS}) \cite{SMW}; this is because
fermions do not couple classically to $\w_\mu$ \cite{Kugo}. Of the SM action 
only the Higgs sector couples to $\tilde R$ (as in  $\tilde R H^\dagger H$)
and also to $\w_\mu$ through its
kinetic term \cite{SMW}; however, these couplings are not changed by transformations
 (\ref{nat}),  (\ref{correspondence}) considered here, hence our results do not
change in the presence of SM. More details will be presented elsewhere \cite{Cezar}.

\section{Conclusions}

We reviewed (non-supersymmetric) gauge theories of $d=4$ space-time symmetries
and studied their quadratic action.
In our view, such gauge theory should:  a) have, as a theory of gravity,
an exact geometric interpretation and origin for their degrees of freedom,
b) recover Einstein gravity in their (spontaneously) broken phase, and c) this symmetry
should be anomaly-free, as any (quantum) gauge symmetry.
Theories based on Weyl gauge group (Poincar\'e $\times$ dilatations)
meet these criteria. However, gauging the full conformal group does not
generate a true gauge theory since  the associated gauge bosons (of special
conformal symmetry and dilatation) are  not physical (dynamical).
In other words, conformal gravity is a gauge theory of conformal
group  as much as, say,  the electroweak theory without  kinetic
terms for $W^\pm$, $Z$ gauge bosons is a gauge theory of weak interactions.

The gauge theory of the Weyl group gives rise to  Weyl quadratic gravity
and this is naturally realised in Weyl conformal geometry where this gauge
  symmetry  is  built in. This quadratic gravity (gauge)
  theory has two equivalent geometric formulations,
that have the same action and thus same physics: a familiar formulation with vectorial
non-metricity but no torsion, and a formulation that is manifestly Weyl-covariant and metric
with respect to a new differential operator ($\hat\nabla$). The theory
recovers Einstein gravity in its (spontaneously) broken phase. In the absence of the SM
all degrees of freedom have geometric origin, and the gauge symmetry is
manifestly maintained in $d$ dimensions which indicates it is anomaly-free,
as it was recently shown elsewhere.

To clarify the origin of the above equivalence,
we compared these two equivalent geometric formulations of Weyl gauge symmetry
to  the standard, modern approach of constructing a gauge theory (of dilatations)
by using the tangent space-time formulation ``uplifted'' to space-time  by the vielbein.
This lead to a gauge theory of dilatations that has an identical associated
quadratic gravity action and that is  metric but has  vectorial torsion.
This third  formulation  is ``dual'' (equivalent) to the non-metric formulation
in  Weyl geometry, to which it is related by a simple projective transformation.
This duality vectorial non-metricity vs vectorial
torsion  was  explained in detail. This equivalence  fails
beyond the vectorial non-metricity and vectorial torsion,  due to the
different, additional  number of degrees of freedom of these tensors in the general case
(that even break the Weyl gauge symmetry of the action). The above three equivalent realisations  of Weyl gauge symmetry: non-metric, Weyl-covariant and
metric with torsion remain equivalent when the SM is added.
The above results suggest that the gauged dilatation may be a fundamental symmetry beyond
both the SM and Einstein gravity and deserves  further investigation.

\bigskip
\noindent
 {\bf Acknowledgements: } This work was supported by project number PN 23210101/2023.
 The work of D.G. was also supported by a grant of  Ministry of Education and Research of
 Romania, CNCS-UEFISCDI project number PN-III-P4-ID-PCE-2020-2255.
 D. Ghilencea  thanks Roberto Percacci (SISSA Trieste),
   Fernando Quevedo (DAMTP, University of Cambridge and CERN TH)
   and Mikhail Shaposhnikov (University of Lausanne), for discussions on
 this topic.

\end{document}